\documentclass[apj]{emulateapj}
\usepackage{apjfonts} 

\usepackage{graphicx}

\newcommand{\Fermi}{{\it{}Fermi}\ }
\newcommand{\GLAST}{{\it{}GLAST}\ }

\makeindex
\citeindextrue

\received{2009 February 12}
\accepted{2009 March 23}
\journalinfo{Astrophysical~journal Letters}
\submitted{}

\shorttitle{Relation between AGN $\gamma$-ray emission and parsec-scale radio jets}
\shortauthors{Kovalev et al.}

\begin{document}
\title{On the relation between AGN gamma-ray emission and parsec-scale radio jets}
\author{
Y. Y. Kovalev\altaffilmark{1,2},
H. D. Aller\altaffilmark{3},
M. F. Aller\altaffilmark{3},
D. C. Homan\altaffilmark{4},
M. Kadler\altaffilmark{5,6,7,8},
K. I. Kellermann\altaffilmark{9},
Yu. A. Kovalev\altaffilmark{2},\\
M. L. Lister\altaffilmark{10},
M. J. McCormick\altaffilmark{4},
A. B. Pushkarev\altaffilmark{1,11,12},
E. Ros\altaffilmark{1,13},
J. A. Zensus\altaffilmark{1,9}
}
\altaffiltext{1}{
Max-Planck-Institut f\"ur Radioastronomie, Auf dem H\"ugel 69, 53121 Bonn, Germany; 
\email{ykovalev,apushkar,ros,azensus@mpifr.de}
}
\altaffiltext{2}{
Astro Space Center of Lebedev Physical Institute,
Profsoyuznaya 84/32, 117997 Moscow, Russia
}
\altaffiltext{3}{
Department of Astronomy, University of Michigan, 817 Denison
Building, Ann Arbor, MI 48109-1042, USA;
\email{haller,mfa@umich.edu}
}
\altaffiltext{4}{
Department of Physics and Astronomy, Denison University,
Granville, OH 43023, USA;
\email{homand,mccorm\_m@denison.edu}
}
\altaffiltext{5}{
Dr.\ Remeis-Sternwarte Bamberg, Universit\"at Erlangen-N\"urnberg,
Sternwartstrasse 7, 96049 Bamberg, Germany
}
\altaffiltext{6}{
Erlangen Centre for Astroparticle Physics, Erwin-Rommel Str.~1,
91058 Erlangen, Germany
\email{matthias.kadler@sternwarte.uni-erlangen.de}
}
\altaffiltext{7}{
CRESST/NASA Goddard Space Flight Center, Greenbelt, MD 20771, USA
}
\altaffiltext{8}{
Universities Space Research Association, 10211
Wincopin Circle, Suite 500 Columbia, MD 21044, USA
}
\altaffiltext{9}{
National Radio Astronomy Observatory, 520 Edgemont Road,
Charlottesville, VA 22903-2475, USA;
\email{kkellerm@nrao.edu}
}
\altaffiltext{10}{
Department of Physics, Purdue University, 525 Northwestern
Avenue, West Lafayette, IN 47907, USA;
\email{mlister@purdue.edu}
}
\altaffiltext{11}{
Pulkovo Observatory, Pulkovskoe Chaussee 65/1, 196140 St. Petersburg, Russia
}
\altaffiltext{12}{
Crimean Astrophysical Observatory, 98409 Nauchny, Crimea, Ukraine
}
\altaffiltext{13}{
Departament d'Astronomia i Astrof\'{\i}sica, Universitat de Val\`encia,
E-46100 Burjassot, Valencia, Spain
}

\begin{abstract}
We have compared the radio emission from a sample of parsec-scale AGN
jets as measured by the VLBA at 15~GHz, with their associated
$\gamma$-ray properties that are reported in the \Fermi LAT 3-month
bright source list. We find in our radio-selected sample that the
$\gamma$-ray photon flux correlates well with the quasi-simultaneously
measured compact radio flux density. The LAT-detected jets in our
radio-selected complete sample generally have higher compact radio flux
densities, and their parsec-scale cores are brighter (i.e., have higher
brightness temperature) than the jets in the LAT non-detected objects.
This suggests that the jets of bright $\gamma$-ray AGN have
preferentially higher Doppler-boosting factors.
In addition, AGN jets tend to be found in a more active radio state within
several months from LAT-detection of their strong $\gamma$-ray emission.
This result becomes more pronounced
for confirmed $\gamma$-ray flaring sources. We identify the parsec-scale
radio core as a likely location for both the $\gamma$-ray and radio
flares, which appear within typical timescales of up to a few months
of each other.
\end{abstract}

\section{INTRODUCTION} 
\label{s:intro}

A number of authors have suggested a close connection between AGN
$\gamma$-ray and radio emission, with the strong energetics of the
jetted relativistic outflow being responsible for accelerating
particles to the high energies needed for $\gamma$-ray production
\citep[e.g.,][]{DS93,SBR94,BB99}. Most of the high energy $\gamma$-ray
sources detected by the EGRET telescope of the {\it Compton Gamma
Ray Observatory} \citep{3EG} were identified with blazars
\cite[e.g.,][]{Mattox_etal01,SRM03,SRM04} which indicates that
relativistic jets with strong Doppler boosting are dominant sites
of extragalactic $\gamma$-ray production.  The EGRET results,
however, were limited by the sensitivity of the telescope and the
erratic sampling of observations. This did not allow a full systematic
comparison of the $\gamma$-ray and radio properties for a complete
sample of blazars.

\cite{Jorstad_etal01} carried out 43~GHz Very Long Baseline Array (VLBA)
monitoring of a sample of EGRET blazars and concluded that $\gamma$-ray
events originate in parsec-scale radio knots that move down the jet at
apparent superluminal speeds. \cite{LV03} supported this finding on the
basis of 37~GHz total flux density single-dish monitoring observations
of EGRET blazars. They concluded that the highest levels of $\gamma$-ray
emission are observed during the initial or peak stages of radio/mm-wave
flares. VLBA observations of a radio-selected sample of bright
extragalactic sources at 15~GHz \citep{2cmPaperIII,2cmPaperIV,LH05} have
shown that jets of EGRET-detected blazars are more compact, contain
faster-moving knots, and are more luminous and more linearly polarized
than non-EGRET blazars. Recent analysis of a large 6~cm VLBA survey
\citep[VIPS,][]{VIPS_EGRET07} has shown that EGRET-detected blazars tend
to have higher brightness temperatures and greater core dominance.
However, a direct correlation between radio and $\gamma$-ray flux
density and luminosity
\citep[e.g.,][]{AAH96,Muecke_etal97,LV03,VIPS_EGRET07} was not found.

The {\it Fermi Gamma-Ray Space Telescope} (previously known as \GLAST)
was successfully launched in June~2008. One of the two instruments on
board the space craft, the Large Area Telescope \citep[LAT,][]{LAT09},
is an imaging, wide-field telescope, covering the energy range from
about 20~MeV to more than 300 GeV. Its characteristics allow for a
quasi-continuous systematic study of the whole $\gamma$-ray sky with
unprecedented sensitivity. The \Fermi LAT team has recently released
results of the first three months of observations (August -- October
2008) in the form of a bright source
list\footnote{http://fermi.gsfc.nasa.gov/ssc/data/access/lat/bright\_src\_list/}
\citep{LATBSL}. In this Letter we compare the $\gamma$-ray and
parsec-scale radio properties of the initial ($>10\sigma$) LAT AGN
detections that are positionally associated with bright radio-loud
blazars \citep{LATBSL,LBAS}. We call this sample throughout the paper
`ELBS' --- Extragalactic LAT Bright Source sample.

Throughout this letter, we use the term ``core'' as the apparent origin
of AGN jets that commonly appears as the brightest feature
in VLBI images of blazars \citep[e.g.,][]{L98,Marscher08}.  We use
the $\Lambda$CDM cosmology with $H_0=71$~km\,s$^{-1}$\,Mpc$^{-1}$,
$\Omega_\mathrm{m}=0.27$, and $\Omega_\Lambda=0.73$
\citep{WMAP5_COSMOLOGY}.

\section{THE RADIO DATA AND SOURCE SAMPLES}
\label{s:obs}

The MOJAVE sample of 135 extragalactic radio sources was rigorously
selected on the basis of the parsec-scale flux density at 15 GHz
($S_\mathrm{VLBA}>1.5$~Jy), and is part of an ongoing monitoring program
with the VLBA at this frequency \citep[for details, see
][]{LH05,Lister_etal09}. The sample consists of the brightest
extragalactic jets north of declination $-20^\circ$, and is dominated by
blazars (101 quasars and 22 BL~Lac objects). Thirty one AGN from this
sample are reported by the \Fermi LAT team in the 3-month \Fermi LAT
Bright $\gamma$-ray Source List \citep{LATBSL}. The MOJAVE program also
began, in late 2008, to monitor weaker LAT-detected AGN (flux density
$S^\mathrm{r}>0.2$~Jy with $\delta > -30^\circ$), some of these were
used for the analysis in \S~\ref{s:gr}, where we supplement the MOJAVE
complete sample with 46 other LAT-detected AGN with
$S^\mathrm{r}>0.2$~Jy. We emphasize that the sources plotted in
Figure~\ref{f:f_corr} do not represent a complete sample selected on
either parsec-scale radio or gamma-ray flux, but this Figure does show
the best available comparison at this time. We also note that many
LAT-detected BL~Lacs \citep{LBAS} are radio weak and do not enter our
radio-selected sample and analysis.

We have attempted to use radio data that are contemporaneous with the LAT
measurements, to avoid, as much as possible, effects due to variability.
We use VLBA observations made during May -- December~2008 for most of the
sources.  For the remaining objects not observed then at the VLBA,
we use recent data from RATAN-600 observations in
September/October 2008 \citep[e.g.,][]{Kovalev_etal99} or the UMRAO
observations in August~2008 -- January 2009 \citep[e.g.,][]{Aller_etal03}.
We infer the 15 GHz parsec-scale flux density with
the assumption that the total single-dish flux well represents the
parsec-scale emission. We used previous RATAN results together with
the VLBA calibrator survey \cite[][and references therein]{vcs6}
to confirm this assumption, following the compactness analysis by
\cite{2cmPaperIV}. Several sources with no VLBA flux density
measurements but with strong kiloparsec-scale emission were not
included in the analysis. If a source was observed more then once
by RATAN or UMRAO, an average value is used.   Note that \cite{2cmPaperIV}
have compared flux density scales of the three programs run at the
VLBA, the UMRAO, and RATAN, and found that they agree within the
errors, that typically are 5~per~cent.

In addition to these quasi-simultaneous measurements of the radio flux
densities, we have used in the analysis in \S~\ref{s:results} (i) the
results from our earlier 15 GHz flux density measurements within the
2cm\,VLBA/MOJAVE program presented by \cite{Lister_etal09}, and (ii)
brightness temperature measurements of the jet cores made since 1999
\citep{2cmPaperIV,Homan_etal06} and recently updated by
\cite{McCormick_AAS09}. We did not restrict the analysis in galactic
latitude, $b$; but galactic plane emission does decrease the LAT
detection sensitivity for $|b|<10^\circ$ \citep{LATBSL}, and this could
bias our conclusions. To test this we re-did our analysis for
$|b|>10^\circ$, and achieved the same results at approximately the same
confidence level in every single test. We also note that the two
brightest radio lenses on the sky which were detected by \Fermi LAT,
B\,0218$+$357 and PKS\,1830$-$211, are not included in the MOJAVE sample
\citep{Lister_etal09} and are not part of the analysis in this paper.

\section{RESULTS}
\label{s:results}
\subsection{Quasi-simultaneous $\gamma$-ray vs.\ radio emission}
\label{s:gr}

\begin{figure}[t]
\begin{center}
\resizebox{0.9\hsize}{!}{
   \includegraphics[trim = 0cm 0.3cm 0cm 0cm]{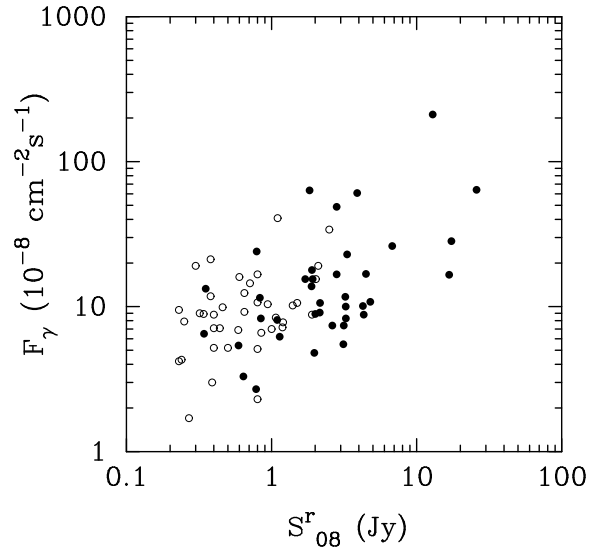}
}
\end{center}
\caption{\label{f:f_corr}
Average \Fermi LAT 100~MeV -- 1~GeV photon flux \citep{LATBSL} versus
quasi-simultaneous 15~GHz flux density.
Filled circles represent total VLBI flux density while open ones~---
single-dish flux density.
The single dish flux densities are representative of the parsec-scale
emission in these objects as described in \S~\ref{s:obs}.
}
\end{figure}

Figure~\ref{f:f_corr} plots the average photon flux measured by the
LAT \citep{LATBSL} in the energy range 100~MeV -- 1~GeV during three months in
August--October, 2008, against the parsec-scale radio flux density,
$S^\mathrm{r}_\mathrm{08}$, measured by the VLBA at 15~GHz within
several months of the LAT observations, namely between May~2008 and
December~2008. This is calculated as the total flux density
over the entire VLBA image. VLBA data points for 38 sources are
supplemented by single-dish measurements for 39 objects as discussed
in \S~\ref{s:obs}. The points in Figure~\ref{f:f_corr} represent the AGN in the
ELBS list ($>10\sigma$ LAT detections) with radio flux density
greater than 0.2~Jy and declination $\delta>-30^\circ$. This sample
is about two times larger than the ELBS-MOJAVE sample. We note that
the redshift distribution of the latter is not statistically different
from that of the full ELBS sample \citep{FM1}.

The non-parametric Kendall's tau test confirms a positive correlation in
Figure~\ref{f:f_corr} at a confidence level $>99.9$~per~cent for the
100~MeV -- 1~GeV energy band.
The test provides the same result when restricted to only those sources
measured by the VLBA (confidence 99.5~per~cent) or when restricted to
the radio brightest MOJAVE-ELBS sources (confidence 99.1~per~cent) or
when restricted to ELBS quasars only (confidence 98.9~per~cent). See
discussion on the possible bias by BL~Lacs in \cite{LBAS}.
The same analysis for the second LAT energy
band, 1--100~GeV, also shows a positive correlation, but at a lower
confidence level, 86~per~cent.

These results indicate a direct relation between the $\gamma$-ray and
parsec-scale synchrotron radiation. We note that \cite{LBAS} found a
much less significant correlation using non-simultaneous total radio
(VLA) flux density in their $\gamma$-ray selected sample. In terms of
previous EGRET results, it is possible that a positive correlation was
not found by, e.g., \cite{AAH96,LV03,VIPS_EGRET07} in the EGRET vs.\
radio analyses because of the lack of homogeneous statistics, limited
EGRET sensitivity and/or use of non-simultaneous measurements in the two
bands. A more thorough analysis with direct implications for the
emission models will be possible when the LAT $\gamma$-ray energy flux
data become available.

\subsection{15 GHz parsec-scale flux density and its variability}
\begin{figure}[t]
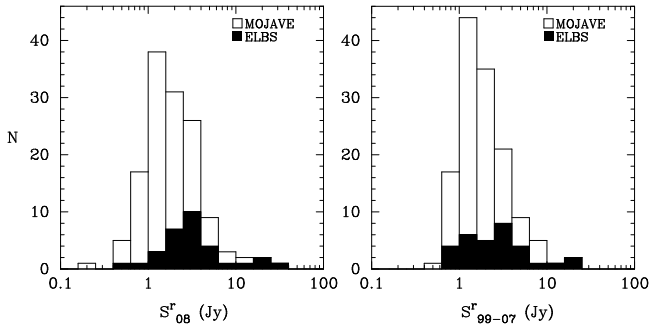

\begin{center}
\resizebox{\hsize}{!}{
   \includegraphics[trim = 0cm 0cm 0cm 0cm]{fig2a.eps}
   \includegraphics[trim = 0.7cm 0cm 0cm 0cm,clip]{fig2b.eps}
}
\end{center}
\caption{\label{f:rflux}
Radio 15~GHz flux density distributions for the complete MOJAVE sample.
{\it Left panel}: mean flux density, $S^\mathrm{r}_{08}$,
measured between May and December 2008. 
{\it Right panel:} mean flux density, $S^\mathrm{r}_{99-07}$,
measured between January 1999 and September~2007. 
The shaded areas show LAT-detected (ELBS) sources. 
}
\end{figure}

The correlation reported in \S~\ref{s:gr} is further supported by the
comparison of the average radio flux density of the objects detected by
\Fermi to that of the non-detected sources for the complete MOJAVE
sample. We assume that LAT non-detected objects have statistically lower
photon flux and therefore should have lower radio flux. We analyzed this
with 114 VLBA data points and 21 single-dish measurements made in 2008
for the complete MOJAVE sample. The $S^\mathrm{r}_\mathrm{08}$ radio
flux densities of parsec-scale jets detected by \Fermi (ELBS sources)
are higher (mean value 3.0~Jy vs.\ 1.6~Jy) and have a different
distribution from the non-detected ones as indicated by the
Kolmogorov-Smirnov (K-S) test at a confidence level 99.9~per~cent
(Figure~\ref{f:rflux}).

\begin{figure}[t]
\begin{center}
\resizebox{0.8\hsize}{!}{
   \includegraphics[trim = 0cm 0.4cm 0cm 0cm]{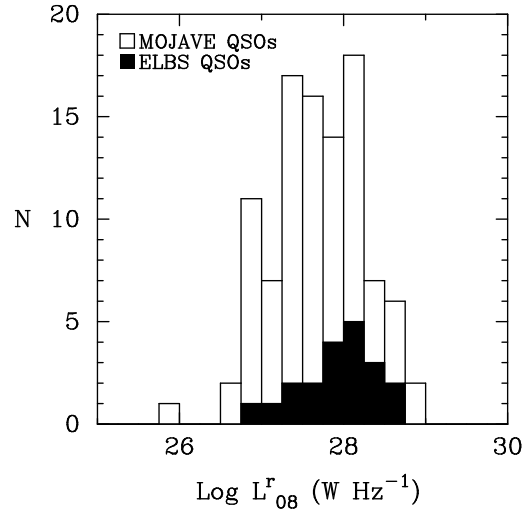}
}
\end{center}
\caption{\label{f:L08q}
Distribution of 15~GHz luminosity calculated from $S^\mathrm{r}_{08}$
(Figure~\ref{f:rvar}) for the quasars in the MOJAVE sample.
The shaded areas indicate LAT-detected (ELBS) quasars.
}
\end{figure}

The MOJAVE
database\footnote{http://www.physics.purdue.edu/astro/MOJAVE/} contains
a complete set of all calibrated VLBA observations of the MOJAVE sources
since 1999 and includes data which we reduced from the public NRAO
archive in addition to our own observations \citep{Lister_etal09}. This
provides us with an estimate of the mean flux density state of each
source, $S^\mathrm{r}_\mathrm{99-07}$, over the previous eight years,
between January~1999 and September~2007. We have made the same
statistical test as above, flux of LAT-detected vs.\ LAT non-detected,
using $S^\mathrm{r}_\mathrm{99-07}$. The general conclusion remains the
same as indicated by the K-S confidence 99.5~per~cent, although the
difference between ELBS and non-ELBS sources becomes less pronounced.
The difference is also found at the 99.6~per~cent K-S confidence for
distributions of the 2008 radio luminosity, $L^\mathrm{r}_{08}$, if
compared for MOJAVE quasars only (Figure~\ref{f:L08q}). BL~Lacs in the
MOJAVE sample have statistically lower redshifts \citep{Lister_etal09}
and were excluded from the luminosity analysis.


\begin{figure}[b]
\begin{center}
\resizebox{0.8\hsize}{!}{
   \includegraphics[trim = 0cm 0.4cm 0cm 0cm]{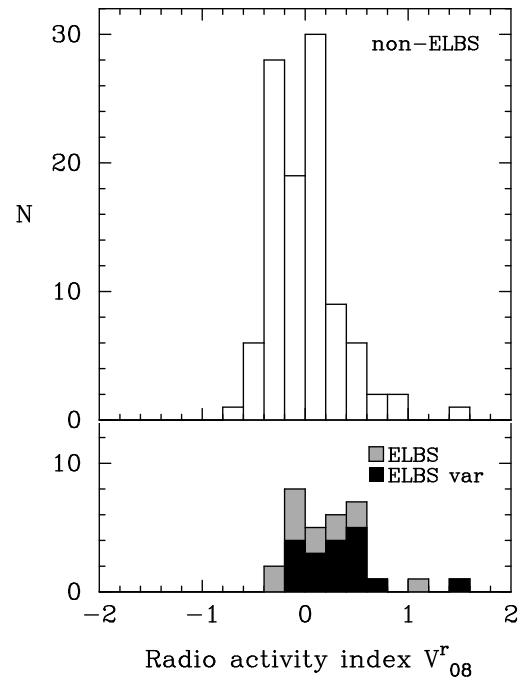}
}
\end{center}
\caption{\label{f:rvar}
Distribution of the radio activity index $V^\mathrm{r}_\mathrm{2008}$
(see definition in the text) for the complete MOJAVE sample. 
The top histogram represents \Fermi LAT non-detected (non-ELBS) sources within
the sample. The bottom histogram represents the
\Fermi AGN $\gamma$-ray detections 
where black shading is used for objects with the 
$\gamma$-ray variability flag from \cite{LATBSL}.
}
\end{figure}

Finally, an intriguing result emerges if we examine the radio
activity level of the ELBS sources by comparing the flux density
measurements in 2008, $S^\mathrm{r}_\mathrm{08}$, with the
respective averages over previous eight years,
$S^\mathrm{r}_\mathrm{99-07}$. We define the 2008 parsec-scale
activity index as
$V^\mathrm{r}_{08}=(S^\mathrm{r}_\mathrm{08}-S^\mathrm{r}_\mathrm{99-07})/S^\mathrm{r}_\mathrm{99-07}$.
The ELBS list includes a flag that indicates whether or not the
source displays significant $\gamma$-ray variability during the
first 3 months of LAT observations \citep{LATBSL}. In
Figure~\ref{f:rvar}, we plot the distributions of our radio
activity index $V^\mathrm{r}_{08}$ for three classes of sources,
based on their LAT detection and LAT variability flag. The
distributions indicate that radio-selected blazars were more likely
to be detected in $\gamma$-rays by the LAT if they were in a
particularly active radio state in May -- December, 2008. 

The average radio activity index $V^\mathrm{r}_{08}$ for the non-ELBS
MOJAVE sources is $-0.01\pm0.03$. For all ELBS-MOJAVE sources it is
$0.23\pm0.07$, and it rises to $0.31\pm0.09$ if ELBS-MOJAVE sources with
the LAT variability flag only are selected. The K-S test confirms, at a
99.0~per~cent confidence, that the ELBS and non-ELBS MOJAVE sources have
different distributions of $V^\mathrm{r}_{08}$. The confidence level
rises to 99.7~per~cent if $V^\mathrm{r}_{08}$ distribution for confirmed
$\gamma$-ray variabile ELBS objects are compared to non-ELBS.
Our conclusions stay the same even if a few outliers with very high
radio activity index $V^\mathrm{r}_{08}>1.0$ are excluded from the
analysis. In this case, the average $V^\mathrm{r}_{08}$ values decrease
to $0.16\pm0.05$ (ELBS-MOJAVE) and $0.24\pm0.07$ (ELBS-MOJAVE with the
$\gamma$-ray variability flag). And we find the following confidence
level of the K-S test for the $V^\mathrm{r}_{08}$ distribution:
97.2~per~cent for ELBS vs.\ non-ELBS, 99.5~per~cent for ELBS confirmed
variable vs.\ non-ELBS. 
The fact that the average value of $V^\mathrm{r}_{08}$ for the entire
MOJAVE sample ($0.05\pm0.03$) is close to zero demonstrates a lack of
overall bias.
It is important to note that the same $V^\mathrm{r}$ activity analysis
was performed for VLBA data prior to 2008 and has shown no significant
difference between ELBS and non-ELBS MOJAVE sources.

Thus we observe a radio activity index significantly greater than zero
for the ELBS sources {\em only} for the time interval of VLBA
observations which roughly coincides with the ELBS detections.
This temporal coincidence suggests that the $\gamma$-rays are produced
in a region close to or coincident with the radio core where radio
flares originate \citep[e.g.,][]{2cmPaperIV}.
However, more densely sampled $\gamma$-ray \Fermi LAT --- radio VLBI
and/or single dish monitoring is necessary to establish this robustly as
the apparent time compression in relativistic jets can be quite large.
The frequency dependent opacity effect which delays radio flares should
also be accounted for.

Our conclusion is supported by the fact that we find extremely few
blazars in their low radio activity state to be $\gamma$-ray bright.
It is also interesting to note that the ELBS sources in Figure~\ref{f:rvar}
with no variability flag (grey shading) span both moderate and high
$V^\mathrm{r}_{08}$ states. This suggests that AGN in this category
might consist of both true $\gamma$-ray non-variable objects as
well as moderately flaring ones.

\subsection{Brightness temperature of jet cores}

\begin{figure}[b]
\begin{center}
\resizebox{0.8\hsize}{!}{
   \includegraphics[trim = 0cm 0.4cm 0cm 0cm]{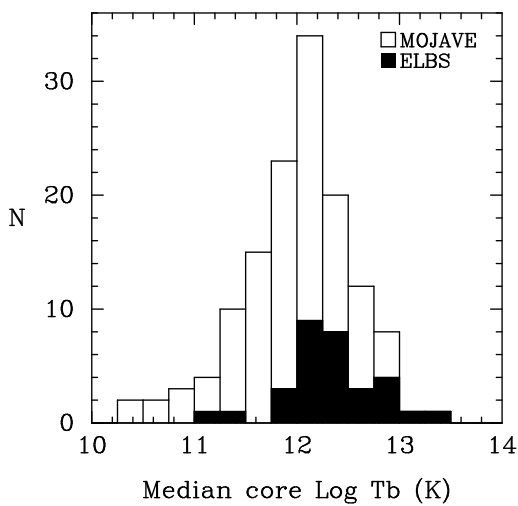}
}
\end{center}
\caption{\label{f:Tb}
Distribution of median brightness temperature values, $T_\mathrm{b}$ of
VLBI cores in the complete MOJAVE sample. The shaded areas represent the
LAT-detected objects in the sample.
}
\end{figure}

Further insight into the $\gamma$-ray/radio jet connection can be
obtained by examining the radio core brightness temperature $T_\mathrm{b}$.
We have calculated median values of $T_\mathrm{b}$ for the MOJAVE sample
for the period 1999-2007 \citep{McCormick_AAS09}. 
In less than 20~per~cent of the cases only a lower limit could be found;
this group of sources includes two ELBS objects.
Gehan's generalized
Wilcoxon test from the ASURV survival analysis package \citep{ASURV}
indicates, at a 99.9~per~cent confidence, that the median $T_\mathrm{b}$
values for LAT-detected sources are statistically higher
than those for the rest of the sample (Figure~\ref{f:Tb}). The
same result is found if the maximum $T_\mathrm{b}$
values are used in the analysis. Based on earlier brightness
temperature findings of \cite{Homan_etal06}, these results suggest
higher Doppler factors for the LAT-detected sources, although their
recent work \citep{McCormick_AAS09} indicates that variations in the
intrinsic $T_\mathrm{b}$ among jets may also play an
important role in the brightest objects. Therefore, we expect
that they might also have faster apparent jet speeds, which is
indeed confirmed by results of the direct kinematics analysis
presented by \cite{FM1}. Lister et al.\ have shown that LAT-detected quasars
in the complete MOJAVE sample have preferentially faster jet motions
than the non-detected ones.

\section{SUMMARY}
\label{s:sum}

On the basis of the joint analysis of the \Fermi $\gamma$-ray LAT and
radio observations of parsec-scale jets in blazars we conclude the
following. The $\gamma$-ray and parsec-scale radio emissions are
strongly related in bright $\gamma$-ray objects detected by \Fermi. At
radio wavelengths, $\gamma$-ray bright sources are found to be
preferentially brighter and more compact, which suggests that they might
have higher Doppler factors than other blazars. The correlations found
suggest that the prominent flares in both $\gamma$-ray and radio bands
are produced in the cores of parsec-scale jets, typically within an
apparent time separation of up to a few months. These findings could be
a consequence of relativistic beaming that boosts the jet emission in
both bands.

The first three months of \Fermi observations represent a significant
improvement over the earlier EGRET results, due to the dramatic increase
in sensitivity and temporal coverage. The combination of these \Fermi
LAT measurements and the extensive VLBA monitoring at 15~GHz by the
MOJAVE program has proved to be a powerful tool to study the nature of
the emission processes in extragalactic jets. We found a clear
connection between the beamed relativistic radio jets and the
$\gamma$-ray emission expected to originate in these regions, confirming
and enhancing earlier results obtained by comparing radio data with the
EGRET catalog.

\acknowledgments
We thank M.~H.~Cohen and A.~P.~Lobanov for thorough reading of the
manuscript and fruitful discussions. 
The authors wish to acknowledge the contributions of the rest of the 
MOJAVE team as well as students at
Max-Planck-Institut f\"ur Radioastronomie and Purdue University.
We also thank J.~McEnery, D.~Thompson and the \Fermi LAT team for discussions of
their plans for publishing their bright source list and AGN list, and we
look forward to future cooperation with the LAT team.
This research has made use of data from the MOJAVE database that is
maintained by the MOJAVE team \citep{Lister_etal09}.
The MOJAVE project is supported under National Science Foundation grant
AST-0807860 and NASA \Fermi grant NNX08AV67G.
D.~C.~H.\ and M.~J.~M.\ were supported by NSF grant AST-0707693.
Part of this project was done while Y.~Y.~K.\
was working as a research fellow of the Alexander von~Humboldt Foundation.
\facility[NRAO(VLBA)]{The VLBA is a facility of the National Science
Foundation operated by the National Radio Astronomy Observatory under
cooperative agreement with Associated Universities, Inc.}
\facility[UMRAO]{
UMRAO is supported in part by a series of grants from the NSF, most
recently AST-0607523, and by funds from the University of Michigan.}
\facility[RATAN]{\mbox{RATAN-600} observations were partly supported by
the NASA JURRISS Program (project W-19611), and the Russian Foundation
for Basic Research (projects 01-02-16812, 05-02-17377, 08-02-00545).}
This research has made use of the NASA/IPAC Extragalactic Database (NED)
which is operated by the Jet Propulsion Laboratory, California Institute
of Technology, under contract with the National Aeronautics and Space
Administration. This research has made use of NASA's Astrophysics Data
System.

{\it Facilities:} \facility{VLBA, UMRAO, RATAN, \Fermi(LAT)}.


\end{document}